\documentclass[aps,prb,twocolumn,superscriptaddress]{revtex4}
\pdfoutput=1
\usepackage{graphicx}
\usepackage{array}
\begin{document}

% Use the \preprint command to place your local institutional report
% number in the upper righthand corner of the title page in preprint mode.
% Multiple \preprint commands are allowed.
% Use the 'preprintnumbers' class option to override journal defaults
% to display numbers if necessary
%\preprint{}
%Title of paper
\title{Doping effects of Cr on the physical properties of BaFe$_{1.9-x}$Ni$_{0.1}$Cr$_{x}$As$_{2}$}

\author{Dongliang Gong}
\affiliation{Beijing National Laboratory for Condensed Matter
Physics, Institute of Physics, Chinese Academy of Sciences, Beijing
100190, China}
\affiliation{University of Chinese Academy of Sciences, Beijing 100049, China}
\author{Tao Xie}
\affiliation{Beijing National Laboratory for Condensed Matter
Physics, Institute of Physics, Chinese Academy of Sciences, Beijing
100190, China}
\affiliation{University of Chinese Academy of Sciences, Beijing 100049, China}
\author{Rui Zhang}
\affiliation{Department of Physics and Astronomy, Rice University, Houston, Texas 77005, USA}
\author{Jonas Birk}
\affiliation{Laboratory for Neutron Scattering and Imaging, Paul Scherrer Institute, CH-5232 Villigen, Switzerland}
\author{Christof Niedermayer}
\affiliation{Laboratory for Neutron Scattering and Imaging, Paul Scherrer Institute, CH-5232 Villigen, Switzerland}
\author{Fei Han}
\affiliation{Department of Nuclear Science and Engineering and Department of Materials Science and Engineering, Massachusetts Institute of Technology, Cambridge, MA 02139, USA}
\author{S. H. Lapidus}
\affiliation{Materials Science Division, Argonne National Laboratory, Argonne, Illinois 60439, USA}
\author{Pengcheng Dai}
\affiliation{Department of Physics and Astronomy, Rice University, Houston, Texas 77005, USA}
\affiliation{Department of Physics, Beijing Normal University, Beijing 100875, China}
\author{Shiliang Li}
\affiliation{Beijing National Laboratory for Condensed Matter
Physics, Institute of Physics, Chinese Academy of Sciences, Beijing
100190, China}
\affiliation{University of Chinese Academy of Sciences, Beijing 100049, China}
\affiliation{Collaborative Innovation Center of Quantum Matter, Beijing, China}
\author{Huiqian Luo}
\email{hqluo@iphy.ac.cn}
\affiliation{Beijing National Laboratory for Condensed Matter
Physics, Institute of Physics, Chinese Academy of Sciences, Beijing
100190, China}

%\maketitle must follow title, authors, abstract, \pacs, and \keywords
\begin{abstract}
We present a systematic study on the heavily Cr doped iron pnictides BaFe$_{1.9-x}$Ni$_{0.1}$Cr$_{x}$As$_{2}$ by using elastic neutron scattering, high-resolution synchrotron X-ray diffraction (XRD), resistivity and Hall transport measurements. When the Cr concentration increases from $x=$ 0 to 0.8, neutron diffraction experiments suggest that the collinear antiferromagnetism persists in the whole doping range, where the N\'{e}el temperature $T_N$ coincides with the tetragonal-to-orthorhombic structural transition temperature $T_s$, and both of them keeps around 35 K. The magnetic ordered moment, on the other hand, increases within increasing $x$ until $x=$ 0.5, and then decreases with further increasing $x$. Detailed refinement of the powder XRD patterns reveals that the Cr substitutions actually stretch the FeAs$_4$ tetrahedron along the $c-$axis and lift the arsenic height away Fe-Fe plane. Transport results indicate that the charge carriers become more localized upon Cr doping, then changes from electron-type to hole-type around $x=$ 0.5. Our results suggest that the ordered moment and the ordered temperature of static magnetism in iron pnictides can be decoupled and tuned separately by chemical doping.

\end{abstract}


\maketitle

\section{Introduction}

Unconventional superconductivity (SC) emerges in the iron pnictides after suppressing the three-dimensional antiferromagnetic (AF) order by chemical doping or high pressure \cite{kamihara2008,hosono2015,stewart2011,chen2014,si2016}. Understanding the nature of magnetism is therefore one of the most important issues to reveal the mechanism of superconductivity in these fascinating materials. In one of the typical parent compounds, BaFe$_2$As$_2$, a stripe type AF order with co-linear structure is formed, where the magnetic moments are aligned along orthorhombic $a$-axes at low temperature (Fig.1(c)) \cite{pdai2012,inosov2016,pdai2015,surmach2017}. For the electron doped BaFe$_{2-x}$$TM_x$As$_2$ ($TM$ = Ni or Co) system, the long-ranged AF  order changes to a short-ranged incommensurate magnetic order and disappears at a finite temperature just before reaching the optimal superconductivity \cite{rotter2008,sefat2008,ljli2009,qhuang2008,chu2009,nni2010,pratt2011,hqluo2012,kim2012,xylu2013,xylu2014,dioguardi2013}. For the hole-doped Ba$_{1-x}$$A_x$Fe$_{2}$As$_2$ ($A$ = K or Na) system, a new type AF order with C$_4$ rotation symmetry interrupts the stripe magnetism in underdoped regime \cite{rotter2008,hqluo2008,avci2012,avci2014,bohmer2015}. For the iso-valently doped BaFe$_{2}$(As$_{1-x}$P$_x$)$_2$ system with less impurity effects, the long-ranged AF order also vanishes in a weakly first-order fashion near the optimal doping \cite{jiang2009,shibauchi2014,allred2014,dhu2015}, similar to the electron doped case. Moreover, superconductivity also emerges in 4$d$ and 5$d$ metal (e.g. Ru, Rh, Pd, Ir, Pt) doped compounds \cite{thaler2010,nni2009,han2009,kirshenbaum2010}. However, the substitution of iron by other transition metals like Cu, Cr and Mn into BaFe$_2$As$_2$, does not induce any superconductivity but only suppress the collinear AF order \cite{nni2010,kim2012,canfield2009,kim2015,sefat2009,marty2011,kim2010,thaler2011,kim2011}. Particularly in Cr and Mn doped compounds, the ordered moments align along the $c-$axes with a checkerboard pattern for heavily doping levels, the so called G-type AF order \cite{pdai2015,kim2012,marty2011,kim2010,kim2011}. For iron pnictides with a superconducting dome, the magnetically ordered temperature (N\'{e}el temperature) $T_N$ is strongly associated with the ordered moments $M$. Both of them are suppressed upon doping, where $M$ is reduced further upon entering the superconducting state, and finally disappears around the optimal doping with finite $T_N$ above the superconducting transition temperature ($T_c$), resulting in an avoided quantum critical point (QCP) due to strong competition between SC and AF  order \cite{si2016,pdai2015}. Similar behavior is also found in the electron doped NaFe$_{1-x}$$TM_x$As ($TM$ = Co, Ni, Cu) system \cite{afwang2012,gtan2013,gtan2016,gtan2017}. To understand the underlying physics of the magnetism as well as the phase diagram in the iron pnictides, one feasible way is to tune $T_N$ and $M$ separately without the influence from superconductivity. Indeed, in the non-superconducting BaFe$_{2-x}$Cr$_x$As$_2$ system, low Cr dopings only suppress $T_N$ in collinear AF order but keep $M$ almost the same as the parent compound until entering the G-type AF regime \cite{marty2011}.

\begin{figure}[h]
\includegraphics[width=0.47\textwidth]{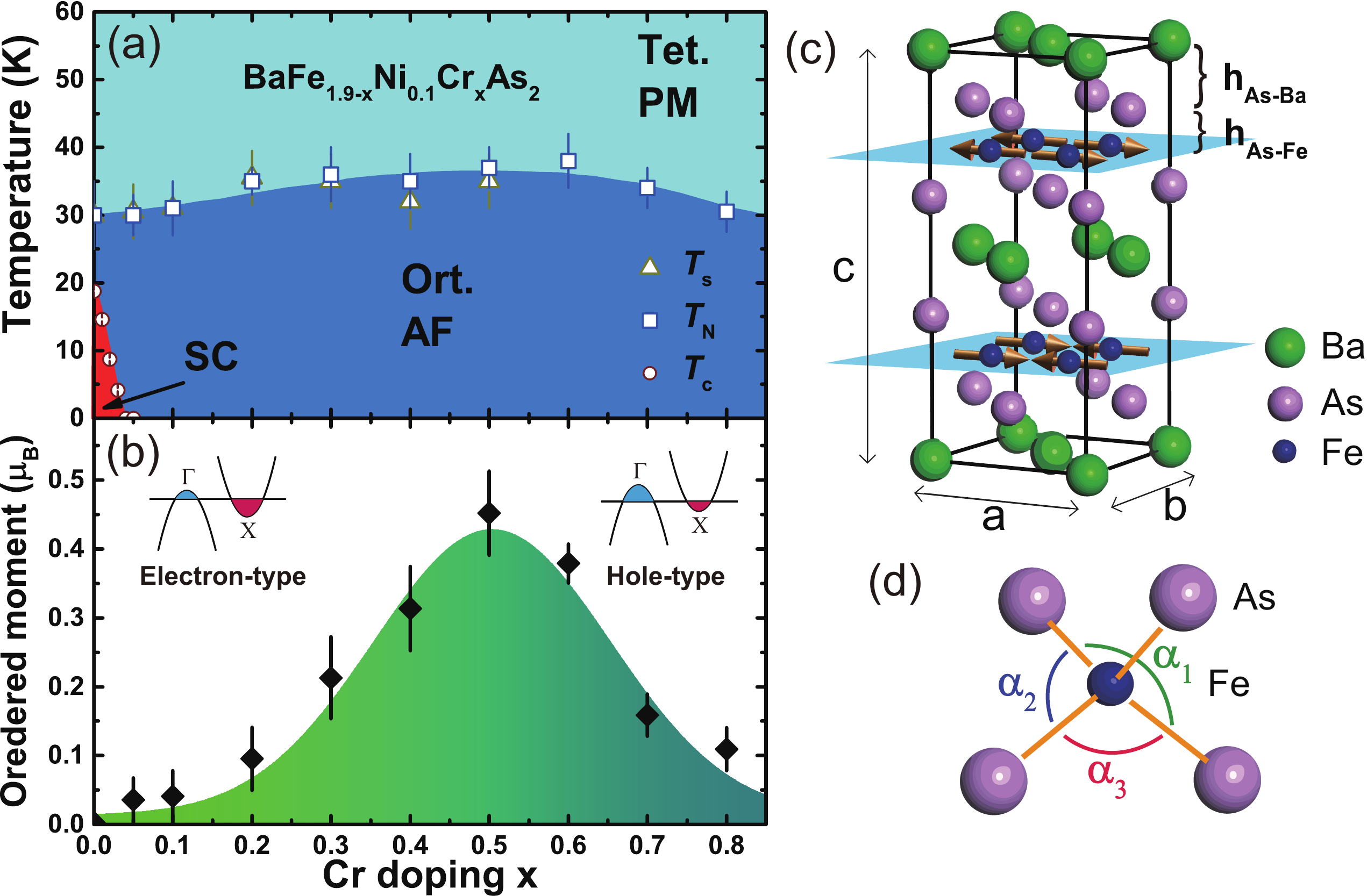}
\caption{(Color online) (a) Phase diagram for BaFe$_{1.9-x}$Ni$_{0.1}$Cr$_{x}$As$_{2}$ determined by neutron diffraction experiments. The PM Tet. and AF Ort. are paramagnetic tetragonal and antiferromagnetic orthorhombic phases, respectively. $T_c$, $T_s$, $T_N$ mark the superconducting transition temperature, structural transition temperature and N\'{e}el temperature, respectively. (b) Cr doping dependence of the effective ordered moment. The insets show schematic band structure and Fermi level below and above $x=0.5$. (c) Crystalline and magnetic structure of BaFe$_2$As$_2$ in orthorhombic phase. (d) Schematic view for As-Fe-As bond angles in iron pnictides.
 }
\end{figure}

In our previous study, we show that Cr is an ideal dopant to suppress superconductivity in BaFe$_{2-x}$Ni$_x$As$_2$ system, where about 1.5\% Cr will totally eliminate the optimal superconductivity in $x=0.1$ compound \cite{rzhang2014}. Only long ranged AF order is found at low temperature in the Cr and Ni co-doped system BaFe$_{2-2x}$Ni$_x$Cr$_x$As$_2$ ($x\leq0.2$) \cite{rzhang2015}. Here, we further push the Cr doping to much higher level than 5\% in the optimal superconducting BaFe$_{1.9}$Ni$_{0.1}$As$_2$ compound, and study the magnetism in a large range of the phase diagram for BaFe$_{1.9-x}$Ni$_{0.1}$Cr$_{x}$As$_2$ ($0 \leq x \leq 0.8$) by neutron diffraction experiments. Unlike the pure Cr doped BaFe$_{2-x}$Cr$_x$As$_2$ system \cite{sefat2009,marty2011}, the collinear AF order persists in the whole explored doping range with quite similar $T_N\approx 35 $ K in BaFe$_{1.9-x}$Ni$_{0.1}$Cr$_{x}$As$_2$, no G-type AF is found until $x=0.8$ (Fig.1(a)). Surprisingly, the ordered moment $M$ strongly depends on the Cr doping, reaching a maximum at $x=0.5$ (Fig.1(b)). Further high resolution X-ray diffraction (XRD) measurements reveal that the Cr substitutions actually stretch the FeAs$_4$ tetrahedron along the $c-$axes and lift the arsenic height (Fig.1(d)). Transport results also reveal that the charge carriers become more localized upon Cr doping, and change from electron-type to hole-type above $x=$ 0.5. Therefore, while the magnetically ordered temperature $T_N$ is mostly determined by the local exchange couplings within the ab-plane, the effective static moments can be tuned by the arsenic height and Fermi surfaces.

\begin{table*}[!htp]
\centering
\caption{ Summary of real composition of BaFe$_{1.9-x}$Ni$_{0.1}$Cr$_{x}$As$_{2}$ single crystal from ICP analysis, the estimation of the correlation length from $\mathbf{Q}$ scans at (1, 0, 1) and (1, 0, 3), the effective ordered magnetic moment for the collinear AF order, the profile factors $R_{p}$, $R_{wp}$ and reduced $\chi^2$ of refinements for all samples at 12 K. The data of $\xi_{ab}$, $\xi_c$ and ordered moment of $x = 0.1$ is extracted from Ref. [\onlinecite{rzhang2015}].}
\begin{tabular}{c|c|p{1.4cm}<{\centering} p{1.4cm}<{\centering}|p{1.5cm}<{\centering}| p{1cm}<{\centering} p{1cm}<{\centering} p{1cm}<{\centering}}
\hline
\hline
$x_{\mathrm{nom}}$ & Ba:Fe:Ni:Cr:As & $\xi_{ab}$ & $\xi_c$ & moment & $R_p$ & $R_{wp}$ & $\chi^2$ \\
 &  & \AA &  \AA & $\mu_B$ &  &  & \\
\hline
%0 &  & 66 & 27  & 0.0027 & 6.25\% & 8.08\% & 2.181 & 0.05542\\
0.05 & 0.98 : 1.86 : 0.08 : 0.04 : 2 & 475(151) & 435(122) & 0.036(32) & 6.43\% & 7.95\% & 1.538 \\
0.1 &  1.00 : 1.82 : 0.08 : 0.06 : 2 & 780(80) & 530(20) & 0.041(37) & 6.09\% & 8.35\% & 2.463 \\
0.2 &  0.98	: 1.76 : 0.08	: 0.14 : 2 & 754(221) & 452(78) & 0.094(46) & 6.11\% & 7.73\% & 1.824 \\
0.3 &  0.98 :	1.68 : 0.08 :	0.20 : 2 & 426(64) & 657(72) & 0.213(59) & 6.74\% & 8.32\% & 1.490 \\
0.4 &  0.98 :	1.60 : 0.08 :	0.26 : 2 & 741(172) & 795(101) & 0.313(61) & 6.42\% & 8.44\% & 2.150 \\
0.5 &  0.98 :	1.55 : 0.08 :	0.34 : 2 & 576(78) & 806(175) & 0.452(61) & 5.8\% & 7.47\% & 1.766 \\
0.6 &  0.98 :	1.50 : 0.08 :	0.42 : 2 & 669(176) & 829(186) & 0.379(28) & 6.67\% & 8.42\% & 2.135 \\
0.7 &  0.98 : 1.40 : 0.08 :	0.50 : 2 & 412(45) & 578(97) & 0.159(31) & 5.58\% & 7.23\% & 1.341 \\
0.8 &  0.98 :	1.36 : 0.08 :	0.56 : 2 & 346(56) & 405(47) & 0.109(31) & 6.77\% & 9.15\% & 2.598 \\
\hline
\hline
\end{tabular}\label{list1.1}
\end{table*}

\section{Experiment details}
High-quality single crystals of BaFe$_{1.9-x}$Ni$_{0.1}$Cr$_{x}$As$_{2}$ were grown by the self-flux method similar to our previous reports \cite{rzhang2014,rzhang2015,ychen2011}. The polycrystalline samples of BaFe$_{1.9-x}$Ni$_{0.1}$Cr$_{x}$As$_{2}$ for XRD experiments were ground from the same batch of the single crystals used in neutron diffraction experiments. The real doping level of Ni and Cr was checked by
inductively coupled plasma (ICP) analysis. The actual and nominal doping levels of both Ni and Cr have linear relationships with the ratios about 0.8 and 0.7 (Table \ref{list1.1}), respectively, consistent with our previous reports \cite{rzhang2014,rzhang2015,ychen2011,dgong2016}. We simply use the nominal composition to represent all samples in this paper for easy comparison with our earlier published results.

XRD measurements of single crystals were carried out on a Mac-Science MXP18A-HF equipment with wavelength of 1.54 \AA{} at room temperature. Synchrotron XRD measurements of polycrystal were performed with wavelength of 0.413 \AA{} at beamline 11-BM-B, Advanced Photon Source (APS), Argonne National Laboratory. All the polycrystalline samples were diluted by the amorphous silicon dioxide (SiO$_{2}$) with ratio of 1:2, and sealed into capillaries with a diameter of 0.3 mm. The capillary was rotated during the measurement to average intensity and reduce the preferred orientation effect. The slew scan range was from -6$^{\circ}$ to 28$^{\circ}$ with a very small step 0.001, which enabled us to obtain the high precision and accuracy data over a 2$\theta$ range from 0$^{\circ}$  to 50$^{\circ}$ based on 12 independent detectors.

Elastic neutron scattering experiments were carried out at the RITA-II cold neutron triple-axis spectrometer at Swiss Spallation Neutron Source, Paul Scherrer Institute, Switzerland. The fixed final energy was $E_f=4.6$ meV with the wavelength of $\lambda_f=4.2$ \AA. To eliminate the scattering from higher-order neutrons with wavelength $\lambda /n (n \geq 2)$, a pyrolytic graphite (PG) filter before the sample and a
cold Be filter after the sample were used. The wave vector $\bf Q$ at ($q_x$, $q_y$, $q_z$) was defined as $(H,K,L) = (q_xa/2\pi, q_yb/2\pi, q_zc/2\pi)$ in reciprocal lattice units (r.l.u.) using the orthorhombic lattice parameters $a \approx 5.61$ \AA, $b \approx 5.59$ \AA{} and $c \approx 13$ \AA. For each doping, a single crystal with a mass of nearly 0.5 grams were aligned to the $[H, 0, 0] \times [0, 0,L]$ scattering plane. The thickness of our sample for neutron scattering was about 0.5 mm, and the neutron absorption was negligible due to small neutron absorption cross sections for all the elements.

The in-plane resistivity ($\rho_{ab}$) was measured by the standard four-probe method with the Physical Property Measurement System (PPMS) from Quantum Design. A large current (5 mA) and slow sweeping rate of temperature (2 K/min) were applied to lower the noise. To compare the temperature dependence of resistivity at different doping concentrations, we normalized the resistivity $\rho_{ab} (T)$ data at room temperature ($T=300$ K). The in-plane Hall resistivity ($\rho_{xy}$) was measured by sweeping the magnetic field at several fixed temperatures. In order to remove the asymmetric electrodes effect and possible magneto-resistance from the sample, the polarity of the magnetic field ($H \parallel c$) was applied from negative to positive during the measurement, where $\rho_{xy}(B)=(\rho_{xy}(+B)-\rho_{xy}(-B))/2$. Further Hall coefficient was obtained from the Hall resistivity after considering the geometry of the sample and electrode contacts. Moreover, the room temperature (300 K) Seebeck coefficient was measured by a homemade system.
\begin{center}
\begin{figure*}[t]
\includegraphics[width=0.95\textwidth]{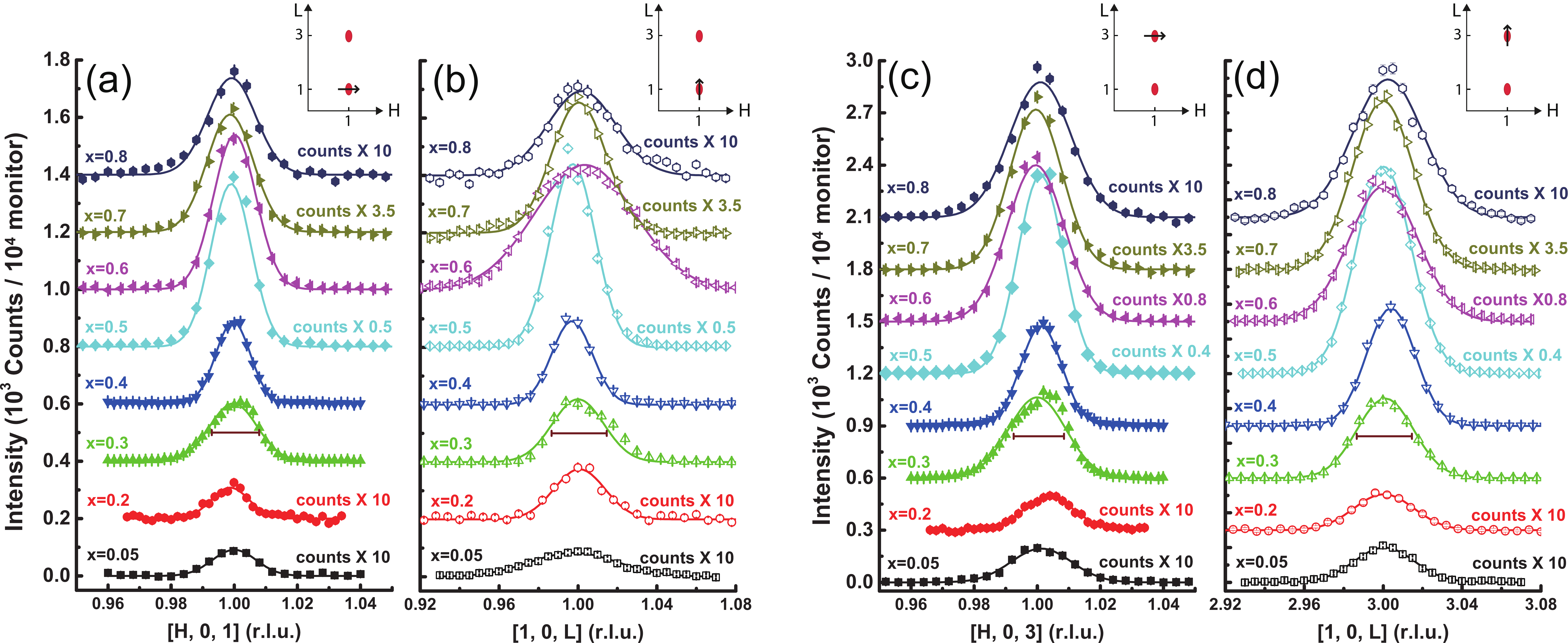}
\caption{(Color online) $\mathbf{Q}$ scans for the antiferromagnetic peaks at
(1, 0, 1) along (a) [H, 0, 1] and (b) [1, 0, L], and (1, 0, 3) along (c) [H, 0, 3] and (d) [1, 0, L] at 2 K. All data are subtracted by the background above $T_N$ and normalized to per $10^4$ monitor counts. For clarity, each peak is multiplied by a ratio shown in the figure and shifted upward in 200 per step for $L=1$, or 300 per step for $L=3$, respectively. The solid lines are Gaussian fitting results. The horizontal bars are the instrumental resolution at $Q=$ (1, 0, 1) and (1, 0, 3) determined by using $\lambda$/2 scattering from the (2, 0, 2) and (2, 0, 6) nuclear Bragg peak above $T_N$ without filter, respectively.
 }
\end{figure*}
\end{center}
\section{Results}

\subsection{Neutron diffraction}

We firstly present our neutron diffraction results. Since the results for $x=$ 0 and 0.1 dopings have been already reported in our previous paper \cite{hqluo2012,xylu2013,rzhang2015}, we won't repeat them here. Figure 2 shows the $\bf Q$ scans for typical collinear AF peaks around $Q=$(1, 0, 1) and  (1, 0, 3) at $T=$ 2 K for $x=$ 0.05, 0.2, 0.3, 0.4, 0.5, 0.6, 0.7, 0.8. Both scan directions along $H$ and $L$ are applied in our experiments. The instrumental resolutions marked as horizontal bars are obtained using $\lambda/2$ scattering from the (2, 0, 2) and (2, 0, 6) nuclear Bragg peak above $T_N$ without filter, e.g. for $\bf Q =$ [H, 0, 1] and [1, 0 ,L] of sample with $x = 0.3$, $R_H$ $\sim$ 0.014 r.l.u. and $R_L$ $\sim$ 0.025 r.l.u., respectively, very similar to our previous experiment at Rita-II spectrometer \cite{hqluo2012,rzhang2015}. All raw data of  Q-scans are subtracted by the flat background above $T_N$, and normalized to per $10^4$ monitor counts. For clarity, each curve is shifted upward by a constant, and multiplied by a ratio to scale with each other as indicated in the figure. We then fit the magnetic Bragg peaks by a flat Gaussian function $I= I_0\exp[-(H-H_0)^2/(2\sigma^2)]$ or $I= I_0\exp[-(L-L_0)^2/(2\sigma^2)]$. The full-width-at-half-maximum (FHWM) $W=2\sqrt{2\ln{2}}\sigma$ is very close to the instrument resolution.  The broadening peak along [1, 0, L] for $x= 0.6$ is due to accidental large sample mosaic.  The spin-spin correlation length $\xi$ in the $ab$ plane and along the $c$ axes are calculated by using the published method \cite{hqluo2012,rzhang2015}, and listed in the Table \ref{list1.1}. For all compositions, both $\xi_{ab}$ and $\xi_c$ are lager than 300 \AA, indicating that the collinear AF order is long ranged for all Cr dopings above $x=0.05$ \cite{marty2011,rzhang2015}.

In elastic neutron scattering, the intensity of the magnetic Bragg peak is proportional to $\mid F_M(Q)\mid^2L$, where $L$ is the Lorentz factor, and $F_M(Q)$ is static magnetic structure factor \cite{gshirane2004}. For a $\bf Q-$scan in triple-axis neutron scattering experiments on single cystal, the Lorentz factor is  determined by \cite{Cowley1988,Longmore1996}:
\begin{equation}\label{eq1}
L^{-1} = \frac{C}{N\sqrt{B_0}}(M_{11}\cos^2\alpha + 2M_{12}\cos\alpha\sin\alpha + M_{22}\sin^2\alpha)^{1/2},
\end{equation}
 where $C$ and $B_0$ depend on instrument parameters, $M_{ij}$ and $N$ are functions of scattering angel $\theta_M$, $\alpha$ is the angle between the wave vector $\bf{Q}$ and the scan direction. For $\theta-2\theta$ scans, we have $\alpha = 0$, thus $L=1/\sin(2\theta_M)$ \cite{Cowley1988}. For $H-$scans in our case (closed to $\theta-2\theta$ scans), the Lorentz factor could be regarded as:
\begin{equation}
L^{-1} = \sin(2\theta_M)R_M.
\end{equation}
Here $R_M$ is a modifying factor from Eq.1 and determined by instrument resolution and scan directions, which is about 0.92 and 1.1 for the $H$ scan at $\bf{Q}$ = (1, 0, 1) and (1, 0, 3), respectively. The magnetic structure factor can be expressed as $F_M(Q)= (\gamma r_0 / 2)\times g\times f_M(Q)\times S\times\sin{\eta}\times\sum(-1)^ie^{iQd}$, where $\gamma$ is the gyromagnetic ratio, $r_0$ is the classical electron radius, $f_M(Q)$ is the magnetic form factor, $S$ is the spin amplitude, and $\eta$ is the angle between the spin $\bf S$ and wave vector $\bf Q$. Therefore, the intensity ratio between $Q$=(1, 0, 1) and (1, 0, 3) is determined by the following equation:
\begin{equation}
\frac{I_{M(101)}}{I_{M(103)}}=\frac{|f_{M(101)}|^2 R_{M(103)}\sin(2\theta_{M(103)})\sin^2\eta_{(101)}}{|f_{M(103)}|^2R_{M(101)} \sin(2\theta_{M(101)})\sin^2\eta_{(103)}}.
\end{equation}
The difference of the form factor between Fe$^{2+}$ and Cr$^{2+}$ is negligible, we have $f_{M(101)}=0.91$ and $f_{M(103)}=0.81$ calculated from their $d-$spacings \cite{hqluo2012,rzhang2015}.
Considering the ordered moment in the collinear antiferromagnetism is along $a-$axes \cite{qhuang2008} and the neutron wavelength $\lambda_f=4.2$ \AA, we have $\theta_{M(101)}=24.2 ^\circ$, $\theta_{M(103)}=37.9 ^\circ$, $\eta_{(101)}=23.3 ^\circ$, $\eta_{(103)}=52.3^\circ$. Then the final intensity ratio is about $I_{M(101)}/I_{M(103)}=0.49$, with 2 times more intensity for the magnetic peak at $Q$=(1, 0, 3) than $Q$=(1, 0, 1) \cite{pdai2015}. Indeed, the statistic of $I_{M(101)}/I_{M(103)}$ from all the raw data in Fig. 2 is about 0.45 $\pm$ 0.04, which is consistent as the expectation within experimental error. This indicates that the collinear antiferromagnetic structure has not been changed in these samples. It should be noticed that the magnetic moments turn to along $c-$axes by forming G-type AF order in the heavily Cr doping BaFe$_{2-x}$Cr$_x$As$_2$ ($x\geq0.6$) with a wavevector ${\bf Q}=[H, H, H]$ in orthorhombic lattice \cite{marty2011,kim2010}. However, in our BaFe$_{1.9-x}$Ni$_{0.1}$Cr$_{x}$As$_{2}$ samples with $x=$ 0.7 and 0.8, we did not find any elastic magnetic scattering at $Q$=(1, 1, 1) by counting 10 times more than $Q$=(1, 0, 1). Perhaps the 5\% Ni doping push away the G-type antiferromagnetism to higher Cr doping in the phase diagram. Due to the mixed signal between magnetic and nuclear scattering at $Q$=(1, 1, 1) in G-type AF order, further polarized neutron scattering experiments are desired to clarify this issue \cite{marty2011}.

Similarly, the intensity of the nuclear Bragg peak is proportional to $\mid F_N(Q)\mid^2L$, where $F_N(Q)$ is the structural factor, and the Lorentz factor $L = 1/\sin(2\theta_N)R_N$ for nuclear scattering angle $\theta_N$ (for $H$ scan around $Q$=(2, 0, 2), $R_N=0.91$). After considering the twinning effect from the magnetic domains, and comparing the intensity of $H$ scans between the magnetic peaks at $Q_{AF}$=(1, 0, 1), (1, 0, 3) at 2 K and nuclear peaks at $Q$=(2, 0, 0), (2, 0, 2) below $T_s$ , we then estimate the static magnetically ordered moment via \cite{hqluo2012,xylu2013,rzhang2015,slli2009}:
\begin{equation}
S=0.067\sqrt{I_MR_M\sin{2\theta_M}/R_NI_N\sin{2\theta_N}}\mid F_N\mid/\mid f_M\mid.
\end{equation}
The error bars of the magnetic moment are determined by the standard deviation of statistics among four case combination of this calculation including the measurement uncertainty of peak intensities. By using the nuclear peak intensities above $T_s$ instead may slightly change the obtained values, but overall the differences are within error bars. Surprisingly, the static moment non-monotonically depends on the Cr doping with a maximum at $x= 0.5$ with 0.452 $\pm$ 0.061 $\mu_B$. For comparison, the cases for $x=$ 0.05 and 0.8 only show small ordered moment with 0.036 $\pm$ 0.032 $\mu_B$ and 0.109 $\pm$ 0.031 $\mu_B$, respectively (Table \ref{list1.1}).

To determine the magnetically ordered temperature ($T_N$), we have measured the temperature dependence of the magnetic Bragg peak intensity at $Q_{AF}$ = (1, 0, 3). The results are shown in Fig. 3(a), where the AF order parameter is proportional to the square root of the magnetic peak intensity, and $T_N$ is defined as the cross point between the linear extrapolations of low temperature AF order parameter and high temperature flat background. For clarity, all data is also normalized to $10^4$ monitor counts and shifted upward in 300 after multiplying the same coefficients in Fig.2(c) and (d). Although the intensity highly depends on the Cr doping due to the variation of ordered moment, the N\'{e}el temperature $T_N$ is almost the same, only ranging from 30 K to 35 K. The orthorhombic-to-tetragonal structural transition temperature $T_s$ can be measured from the temperature dependence of the nuclear peak intensity at $Q$ = (2, 0, 0) due to the neutron extinction release related to the lattice distortion \cite{aigoldman2008,clester2009,jzhao2008,dkpratt2009,akreyssig2010,xylu2014b}, as shown in Fig. 3(b). For Cr doping with $x\leq 0.5$, we have observed coinciding $T_s$ and $T_N$, suggesting strong magneto-elastic coupling in this system \cite{xylu2013,xylu2016}. The huge structural factor and strong intensity of the Bragg peak for higher Cr doping levels makes it difficult to figure out the structural transition via extinction effect. However, due to the breaking symmetry of $O(3) \times Z_2$ in the collinear AF structure, the in-plane rotation symmetry changes from $C_4$ to $C_2$, the lattice must form a orthorhombic phase below $T_N$ to stabilize the magnetism \cite{si2016,pdai2015,dgong2017}.

\begin{figure}[t]
\includegraphics[width=0.45\textwidth]{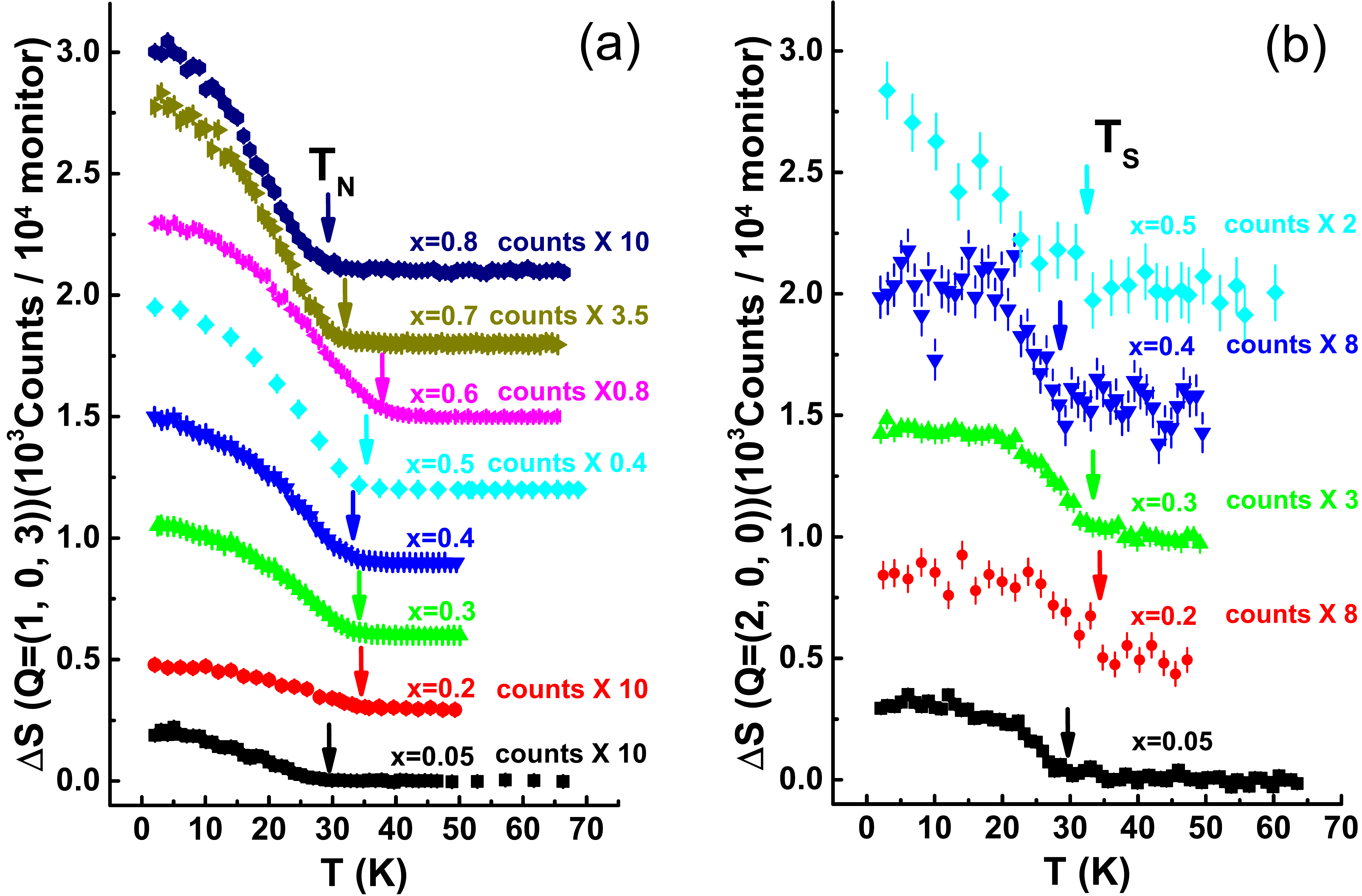}
\caption{(Color online) (a) Temperature dependence of the magnetic peak conuts
at $Q_{AF}$ = (1, 0, 3). All data are subtracted by the background
above $T_N$ and normalized to $10^4$ monitor counts. For clarity, each curve is multiplied by a ratio shown in the figure and shifted upward in 300 per step. (b) Intensity enhancement from neutron extinction release effect at $Q=$ (2, 0, 0). All data are subtracted by the data above $T_s$ and normalized to $10^4$ monitor counts. For clarity, each curve is multiplied by a ratio shown in the figure and shifted upward in 500 per step.
 }\label{fig3}
\end{figure}

By summarizing the structural transition temperature $T_s$, magnetic transition temperature $T_N$ and superconducting transition temperature $T_c$, we plot the phase diagram of the BaFe$_{1.9-x}$Ni$_{0.1}$Cr$_{x}$As$_{2}$ in Fig. 1(a). The initial superconductivity with $T_c=20$ K in optimally doped BaFe$_{1.9}$Ni$_{0.1}$As$_{2}$ is quickly suppressed by Cr doping up to $x=0.04$, accompanying by a recovery of the short-ranged AF order to long-ranged AF order \cite{hqluo2012,rzhang2014,rzhang2015}. However, both $T_N$ and $T_s$ are insensitive to Cr doping except for a slightly enhancement in $x=0.4 \sim 0.6$ compounds, suggesting the robust orthorhombic AF phase in this system. Very different doping dependence between N\'{e}el temperature $T_N$ and effective average moment $M$ is observed in this iron pnictide system (Fig.1(a) and (b)), which means the ordered temperature and ordered strength can be tuned separately by changing the doping ratio of Ni and Cr \cite{chen2014,pdai2015}.

\subsection{X-ray diffraction}

\begin{figure}[t]
\includegraphics[width=0.45\textwidth]{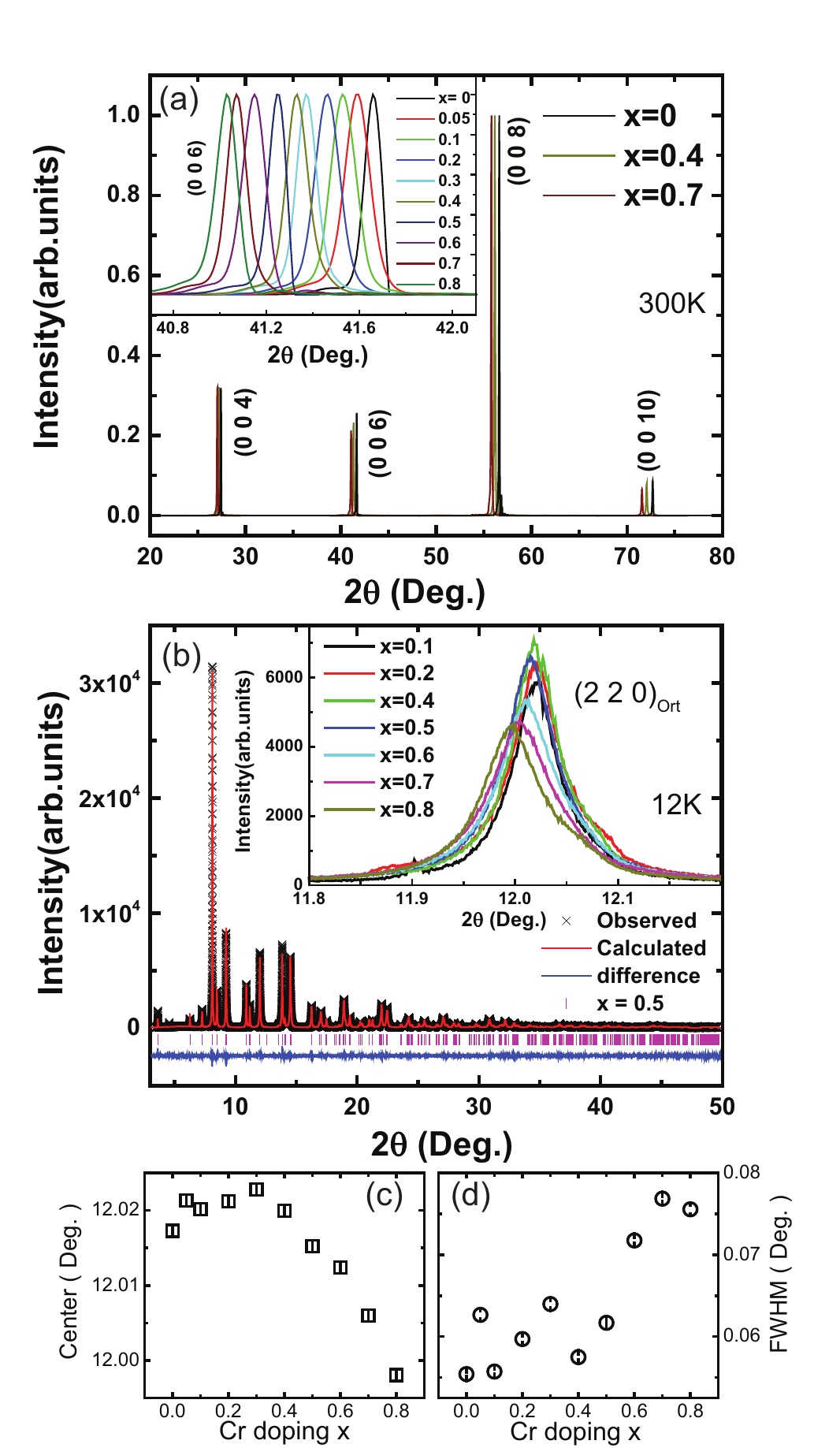}
\caption{(Color online) (a) Typical patterns of X-ray diffraction on the as-grown single crystals BaFe$_{1.9-x}$Ni$_{0.1}$Cr$_{x}$As$_{2}$ with $x$ = 0, 0.4 and 0.7 at room temperature, the inset shows that the peak (0 0 6) shifts with Cr doping increasing. For clarity, the intensity is normalized to [0, 1].(b) Synchrotron X-ray diffraction pattern and refinement results on $x$ = 0.5 powder sample with wavelength of 0.414 \AA{} at 12 K, the inset shows the evolution of Bragg peak (2, 2, 0) upon Cr doping. (c) and (d) Cr doping dependence of the peak center and peak width (FWHM) for (2, 2, 0) nuclear peak.
}
\end{figure}

We have carried out XRD experiments to check the Cr doping effect on the lattice structure. The as-grown single crystal XRD are measured at room temperature with incident beam along $c-$axes, typical results of the normalized data for $x=$ 0, 0.4 and 0.7 are shown in Fig. 4(a), where the insert shows the peak shift of (0, 0, 6) for all samples. The sharp peaks with a narrow width about 0.1$^{\circ}$ indicate the high crystalline quality of our samples. The systematic shift of (0, 0, 6) peak towards to low scattering angle upon Cr doping, suggests that Fe is indeed substituted by Cr, and the lattice is stretched along $c-$axes by larger ionic size of Cr.

To further check the Cr doping effect on the lattice parameters and bonding angles, we have performed high resolution synchrotron XRD on powder samples with wavelength of 0.414 \AA{} at $T=$ 12 K (below $T_s$), 55 K, 270 K (above $T_s$). The polycrystalline samples are ground from the same batch of those single crystals measured in neutron diffraction experiments. Figure 4(b) shows the representative diffraction patterns of $x=0.5$ sample, where the refinement is done by GSAS-EXPGUI package with profile factors $R_p=5.8\%$ and $R_{wp}=7.47\%$. The impurity from the flux Fe$_{1.9-x}$Ni$_{0.1}$Cr$_x$As$_2$ or other phases, even if they exist, should be less than 0.6\% from our refinements. For comparison, we refine the patterns with one phase uniformly for all Cr doping levels, and list the $R_p$, $R_{wp}$ and reduced $\chi ^2$ in the Table \ref{list1.1}. The small values of these parameters concerning the refinement quality indicate that the sample phase is in high purity for such a complex system. The evolution of Bragg peak (2, 2, 0) with different Cr doping at 12 K, which contains the information concerning the $ab$ plane, is shown in the insert of Fig. 4(b). The nearly monotonic change of the peak center and peak width at larger Cr concentrations are found (Fig. 4(c)(d)). Such small shift of the Bragg peak position indicates that the in-plane lattice parameters are slightly affected by Cr doping. The broadening of the peak width may be induced by microstrain from the internal defect by Cr doping rather than the external processing by grinding.

\begin{figure}[t]
\includegraphics[width=0.5\textwidth]{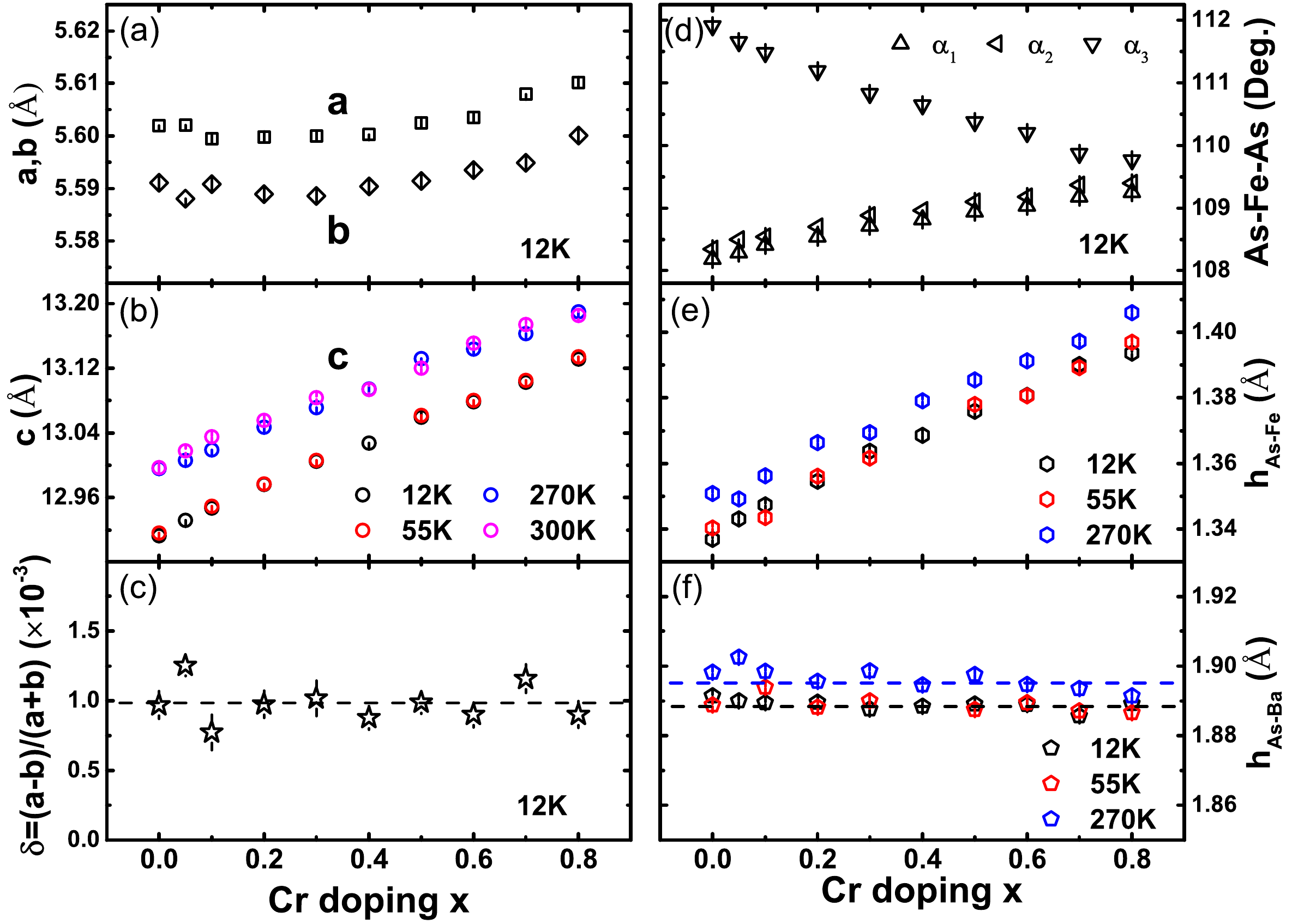}
\caption{(Color online) Cr doping dependence of (a) the lattice parameters $a$ and $b$ at 12 K, (b) the lattice parameter $c$ at 12 K (black), 55 K (red), 270 K (blue) and 300 K (magenta), (c) the lattice distortion $\delta$ at 12 K, (d) As-Fe-As bond angles at 12 K, (e)-(f) arsenic heights to the iron plane and barium plane at 12 K, 55 K, 270 K.
}
\end{figure}

We summarize all refinement results in Fig. 5, including: the lattice parameters $a$, $b$, $c$; the lattice orthorhombicity $\delta$; the bonding angles of As-Fe-As $\alpha_1$, $\alpha_2$, $\alpha_3$; the arsenic heights to the iron-plane $h_{As-Fe}$ and to the barium-plane $h_{As-Ba}$, as defined in Fig. 1(c) and (d). The doping dependence of $c-$axis parameter at 300 K is obtained from the data of single crystal XRD experiments(Fig. 4(a)), which is nearly the same as 270 K data in the powder diffraction experiments. The $c-$axis shrinks about 0.5\% by cooling down to low temperature at 12 K and 55 K due to thermal effects, and continuously increases upon Cr doping (Fig. 5(b)). The in-plane lattice parameters, $a$ and $b$, are weakly dependent on Cr doping with a very small increment after $x=0.5$ (Fig. 5(a)). The lattice orthorhombicity, defined as the in-plane lattice distortion at low temperature: $\delta = (a-b)/(a+b)$, is around 10$^{-3}$ and doping independent within the instrument resolution (Fig. 5(c)). The As-Fe-As bonding angles reach the ideal angle ($109.5^\circ$) of a regular tetrahedron when approaching $x=0.8$ (Fig. 5(d)) \cite{chen2014,hcmao2017}. Therefore, the Cr doping actually stretches the FeAs$_4$ tetrahedron and lifts the arsenic ions away Fe-Fe plane with longer Fe-As distance (Fig.5 (e)). Consequently, the arsenic height to barium-plane would not be affected (Fig. 5(f)), but the $c-$axis is increased simultaneously by increasing Cr doping (Fig.5 (b)). The similar slopes of the doping dependence of $c-$axis and arsenic heights ($h_{As-Fe}$ and $h_{As-Ba}$) below and above $T_s$, prove these behaviors indeed come from Cr doping effect rather than the thermal effect. The lifting of As atoms away from Fe-Fe plane will cause more difficulties for the electron hopping between iron ions via the arsenic intermediary, resulting in a localization effect in the electron transport. Similar effect has been already observed in our previous studies on the BaFe$_{1.7-x}$Ni$_{0.3}$Cr$_{x}$As$_{2}$ system \cite{rzhang2015}.

\subsection{Transport measurements}

\begin{figure}[t]
\includegraphics[width=0.45\textwidth]{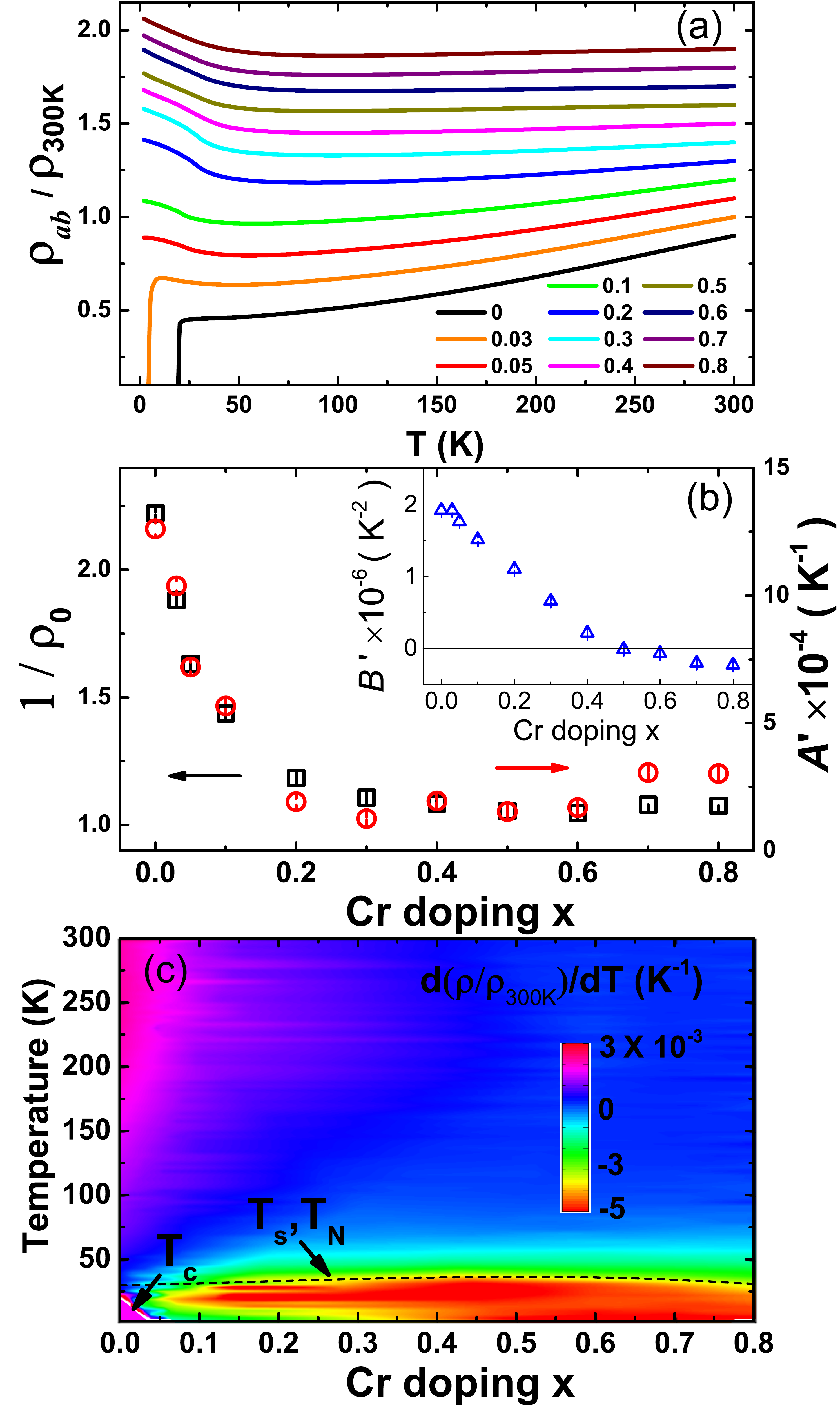}
\caption{(Color online) (a) Temperature dependence of in-plane resistivity for BaFe$_{1.9-x}$Ni$_{0.1}$Cr$_{x}$As$_{2}$.  For clarity, all data are normalized by the data at 300 K and shift upward in 0.1 step for each compound. (b) Doping dependence of the parameters from model fitting results of resistivity between 150 K and 300 K .(c) The gradient color mapping for the temperature and doping dependence of the first order differential of resistivity $d\rho/dT$ of BaFe$_{1.9-x}$Ni$_{0.1}$Cr$_{x}$As$_{2}$, where $T_c$, $T_s$ and $T_N$ mark the superconducting transition temperature, structural transition temperature and N\'{e}el temperature verse Cr doping $x$. The dashed line is obtained from neutron diffraction results in Fig.\ref{fig3}.
 }\label{fig6}
\end{figure}

\begin{figure}[t]
\includegraphics[width=0.45\textwidth]{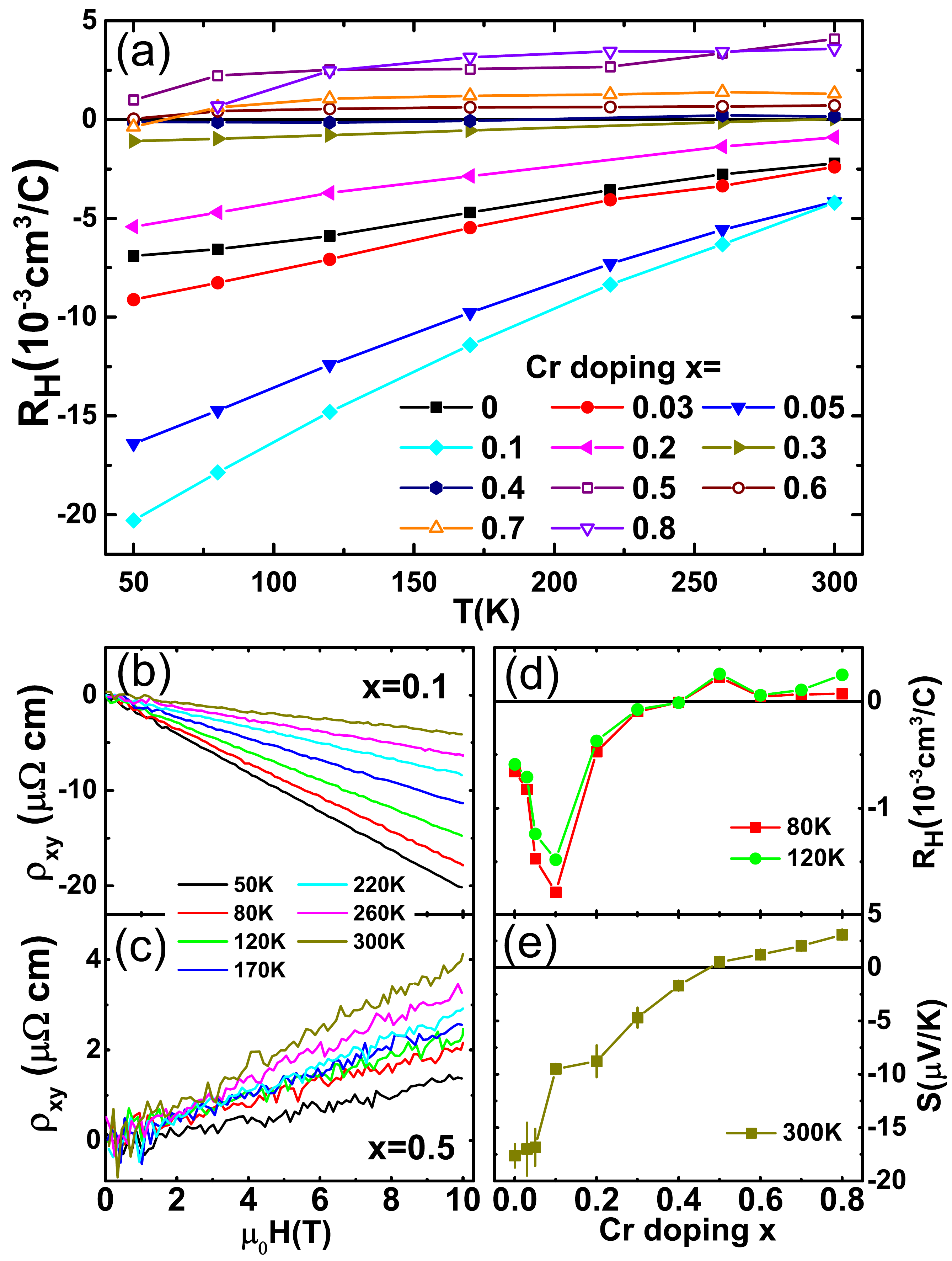}
\caption{(Color online) (a) Temperature dependence of the Hall coefficient $R_H$ for BaFe$_{1.9-x}$Ni$_{0.1}$Cr$_{x}$As$_{2}$. (b) and (c) Magnetic field dependence of the Hall resistivity $\rho_{xy}$ at different temperatures for BaFe$_{1.8}$Ni$_{0.1}$Cr$_{0.1}$As$_{2}$ and BaFe$_{1.4}$Ni$_{0.1}$Cr$_{0.5}$As$_{2}$ , respectively. (d) and (e) Cr doping dependence of the Hall coefficient $R_H$ at 80 K, 120 K and the Seebeck coefficient at 300 K, respectively.
}
\end{figure}

To examine how Cr doping affects on the properties of the charge carriers, we have performed resistivity, Hall coefficient and Seebeck coefficient measurements on BaFe$_{1.9-x}$Ni$_{0.1}$Cr$_{x}$As$_{2}$ system. The normalized resistivity in the $ab$ plane $\rho_{ab}/\rho_{\mathrm{300K}}$ down to 2 K is shown in Fig. 6(a). To understand the behavior of resistivity at the normal state (high temperature), we conduct a fit for the data from 150 K to 300 K by an empirical model: $\rho(T)/\rho_{\mathrm{300K}} = \rho_0 + A^{\prime}T + B^{\prime}T^2$, where $\rho_0$ is the normalized residual resistivity, $A^{\prime}$ is the magnitude of $T$-linear term (so called non-Fermi-liquid term) mostly related to the slope of $\rho(T)$, and $B^{\prime}$ is the magnitude of quadratic term (so called Fermi liquid term).  The fitting parameters are shown in the Fig. \ref{fig6}(b). The decreasing of $1/\rho_0$ and $A^{\prime}$ with increasing $x$ up to 0.5 suggests the metallic behavior is suppressed upon Cr doping due to localization effect of charge carriers. However, the $A^{\prime}$ term slowly recovers above $x= 0.5$, indicating the mobility of the charge carriers may be improved again for higher Cr dopings. The $B^{\prime}$ term switches to a negative value for Cr doping higher than $x=0.5$, which is very similar to the hole doped Ba$_{1-x}$K$_x$Fe$_{2}$As$_2$ system \cite{rotter2008,hqluo2009}. These facts suggest that the Cr doping may affect not only on the mobility but also the type of charge carrier. Except the superconducting compound BaFe$_{1.9}$Ni$_{0.1}$As$_{2}$ with zero Cr doping, all curves in Fig. \ref{fig6}(a) show an upturn at low temperature. This is attributed to the dual effects from charge carrier localization and magnetic transition at low temperature \cite{rzhang2015}, which is very clear from the transition boundary in the gradient color mapping for the first order differential of resistivity $d\rho/dT$ shown in Fig.\ref{fig6}(c).

The effective carrier density can be measured from Hall resistivity at normal state. The temperature dependence of the Hall coefficient $R_H$ above $T_N$ is shown in Fig 7(a). More interestingly, the sign of $R_H$ switches from negative to positive with Cr substitution when $x\geq 0.5$. For clarity, we have also shown the magnetic field dependence of the Hall resistivity for $x$ = 0.1 and 0.5 in Fig. 7(b) and (c), both of them have linear temperature dependence but opposite slope. By selecting the Hall coefficient $R_H$ data at 80 K and 120 K in Fig.7 (d), we find a clear minimum at $x=0.1$ and a sign change between $x=0.4$ and 0.5, even the signal above $x=0.5$ doping is really weak. Therefore, the Cr dopings initially causes localization effect to suppress the superconductivity when $x\leq 0.1$, but then introduce more holes into the system and finally turn the effective charge carriers to be hole-like around $x=0.5$. Similar process occurs in the BaFe$_{2-x}$Cr$_{x}$As$_{2}$ system, where the electron-to-hole crossover is around $x=0.15$ \cite{kobayashi2017}. This speculation is further confirmed by the thermoelectric power measurement, where the Seebeck coefficient also changes sign between $x=$ 0.4 and 0.5 as shown in Fig. 7(e).  Since the effective charge carrier in the parent compound BaFe$_2$As$_2$ already is electron-like \cite{hqluo2009}, the Cr doping, even it would be naively thought to be hole doping, does not have exactly opposite behaviors comparing to the electron doping from Ni due to particle-hole asymmetry \cite{myi2009}. Moreover, the impurity scattering from the ionic substitution may be quite different between Cr and Ni \cite{surmach2017,ideta2013}. In this case of BaFe$_{1.9-x}$Ni$_{0.1}$Cr$_{x}$As$_{2}$, a smeared band structure and hole-like Fermi surfaces are then expected.

\section{Discussion and Conclusions}

It is theoretically predicted that the Fe-pnictogen distance and the shape of FeAs$_4$ tetrahedron have crucial influence on the static moment \cite{yin2011}. In our neutron scattering experiments on the BaFe$_{1.9-x}$Ni$_{0.1}$Cr$_{x}$As$_{2}$ system, the observation of entirely different Cr doping dependences between ordered temperature $T_N$ and effective moment $M$ suggests a complex origin of the magnetism in this iron pnictide system \cite{hirschfeld2011,chubukov2012,basov2011,si2008,fang2008,xu2008,mazin2009}. Since the Cr doping has limited effect on the in-plane lattice parameters, the direct interactions determined by in-plane exchange couplings are nearly unchanged. Due to the strong magneto-elastic coupling in this system \cite{xylu2013}, the nearly doping independent lattice orthorhombicity $\delta$ under weak distortion of in-plane lattice probably makes both $T_s$ and $T_N$ staying around 35 K. However, the Cr substitution strongly stretches FeAs$_4$ tetrahedron by lifting As height, thus the hoppings between Fe-As-Fe indirect interactions become more difficult due to increasing Fe-As distance. In this case, the electron system becomes more localized with enhanced electron correlations and larger effective mass, forming larger static moment upon Cr doping. Meanwhile, the Cr doping actually introduces holes into the system, which compensates the electrons by lowering the chemical potential and reshapes the Fermi surface (Fig. 1(b)). The better condition of Fermi surface nesting stabilizes the magnetic ordering, and the system finally reaches the maximum ordered moment when correlations are strong enough \cite{yin2011,haule2009}. Further Cr doping breaks down the balance between the electron pocket and hole pocket, and switches the effective charge carriers from electron-like to hole-like above $x=0.5$. The mobility of the system will be improved again with itinerary holes and reduced effective mass, which strongly enhances the Fe-Fe direct hopping. The correlation strength may be further enhanced by increasing Fe-As distance, thus less quasiparticles condense to form a reduced static moment in the AF order (Fig.1(b)) \cite{yin2011,hirschfeld2011,chubukov2012,basov2011}. For the highest doping in our studies $x=0.8$, the magnetic moment is still strong along with impurity scattering from Cr, so the superconductivity can not survive for limited hole density even below such low $T_N$ \cite{si2016,pdai2012,chen2014,inosov2016,pdai2015}.

In conclusion, we have systematically studied the antiferromagnetism, crystal structure and electronic transport of the heavily Cr doped BaFe$_{1.9-x}$Ni$_{0.1}$Cr$_{x}$As$_{2}$ system. We find that both magnetically ordered temperature $T_N$ and structural transition temperature $T_s$ keep around 35 K when doping Cr from $x=0.05$ to $x=0.8$, while effective moments are significantly enhanced then suppressed down after $x\geq$ 0.5. Detailed structural analysis suggests that the FeAs$_4$ tetrahedron is stretched by lifting As atoms away Fe-Fe plane but keeping the lattice orthorhombicity unchanged upon Cr doping. A crossover from electron-type to hole-type charge carriers together with their mobility happens around $x=$ 0.5, too. These results suggest that the ordered moment and the ordered temperature of static magnetism in iron pnictides can be tuned separately by different chemical dopings. It seems that the superconductivity occurs more likely in those systems with intermediate correlation strength and sufficient intensity of itinerant electrons or holes.

\begin{center}
{\bf Acknowledgements}
\end{center}
The authors thank the helpful discussion with Zhiping Yin, Daoxin Yao, Wei Ku, Jitae Park and assistance on the thermal power measurements from Huaizhou Zhao.
This work is supported by the National Natural Science Foundation of China (Nos. 11374011, 11374346, 11674406 and 11674372), the Strategic Priority Research
Program (B) of the Chinese Academy of Sciences (XDB07020300), the Key Research Program of the Chinese Academy of Sciences (XDPB01), the National Key Research and Development Program of China (Nos. 2017YFA0303103,2017YFA0302903,2016YFA0300502), and the Youth Innovation Promotion Association of CAS (No. 2016004). Work at Rice is supported by the U.S. NSF-DMR-1700081 and by the Robert A. Welch Foundation Grant No. C-1839 (P.D.).

\end{document}